
\input panda
%
\loadamsmath
%
\def\CP{{{\Bbb C}{\rm P}^{n-1}}}
\def\SU{{{\rm SU}(n)}}
\pageno=0\baselineskip=14pt
\nopagenumbers{
\line{\hfill CERN-TH.7426/94}
\line{\hfill SWAT/93-94/42}
\line{\hfill\tt hep-th/9409142}
\line{\hfill September 1994}
\ifdoublepage \bjump\bjump\bjump\bjump\else\vfill\fi
\centerline{\capsone The exact mass-gap of the supersymmetric $\CP$
sigma model}
\bjump\bjump
\centerline{\scaps Jonathan M. Evans\footnote{$^{*}$}{Supported by a
fellowship from the EU Human Capital and Mobility programme.}
Timothy J.~Hollowood\footnote{$^{**}$}{On leave from:
Department of Physics, University of Wales, Swansea, SA2
8PP, U.K.}}
\sjump
\sjump
\centerline{\sl CERN-TH, CH-1211 Geneva 23, Switzerland.}
\centerline{\tt evansjm@surya11.cern.ch, hollow@surya11.cern.ch}
\bjump\bjump\bjump
\ifdoublepage
\vfill
\noindent
\line{CERN-TH.7426/94\hfill}
\line{September 1994\hfill}
\eject\null\vfill\fi
\centerline{\capsone ABSTRACT}\sjump
A formula for the mass-gap of the supersymmetric $\CP$ sigma
model ($n > 1$) in two dimensions is derived:
$m/\Lambda_{\overline{\rm MS}}=\sin(\pi\Delta)/(\pi\Delta)$ where
$\Delta=1/n$ and $m$ is the mass of the fundamental particle multiplet.
This result is obtained by comparing two expressions for the
free-energy density
in the presence of a coupling to a conserved charge; one expression
is computed from the exact S-matrix of K\"oberle and Kurak via the
thermodynamic Bethe ansatz and the other is computed using conventional
perturbation theory.
These calculations provide a stringent test of the S-matrix, showing
that it correctly reproduces the universal part of the beta-function
and resolving the problem of CDD ambiguities.
\sjump\vfill
\ifdoublepage \else
\noindent
\line{CERN-TH.7426/94\hfill}
\line{September 1994\hfill}\fi
\eject}
\yespagenumbers\pageno=1
%
%
\def\t{\theta}

\chapter{Introduction}

This paper is the second of a pair (see [\Ref{EH}])
concerned with exact results for supersymmetric sigma-models in two dimensions.
The general strategy, which has already been successfully applied to
non-supersymmetric theories in
[\Ref{HN},\Ref{HMN},\Ref{BNNW},\Ref{FNW},\Ref{CGN},\Ref{THIII},\Ref{ACFT}],
is to calculate the free energy of a theory in the presence of a
background field by using the Thermodynamic Bethe Ansatz (TBA)
equations derived from some postulated exact S-matrix.
By comparing this result with a calculation of the same quantity in
standard perturbation theory, the validity of the S-matrix can be checked
and the exact mass-gap for the model can be extracted. In [\Ref{EH}] the
supersymmetric O($N$) sigma model ($N > 4$)
was considered and it was shown how the
novel difficulties associated with a supersymmetric theory---in particular
the problem of diagonalizing the resulting TBA
system---could be overcome for this simplest family of examples.
In this sequel we consider the supersymmetric $\CP$ sigma-models.
These theories differ from the
O($N$) models in having $N=2$ supersymmetry and also in a number
of other important respects.
For our purposes, the most important difference is that the TBA
equations we shall have to solve involve anti-particles as well as particles.
We shall see that the methods described
above can, nevertheless, be applied successfully in these cases too.

The bosonic $\CP$ models ($n>1$) first attracted attention as
``toy'' field theories with instanton solutions [\Ref{Raj}]
generalizing the bosonic O(3) model (the O(3) and ${\Bbb C}$P${}^1$ models
coincide but there are no instanton
solutions in the O($N$) models for $N > 3$).
The analogy with QCD is reinforced by the fact that the bosonic $\CP$
theories are asymptotically free, they generate their mass dynamically,
and they are confining. Unfortunately, the classical
integrability of the bosonic $\CP$ models appears to be vitiated at the
quantum level by anomalies in the conserved currents.
For the supersymmetric $\CP$ models [\Ref{INST},\Ref{DDL2},\Ref{CS}]
on the other hand, integrability is
maintained at the quantum level due to a cancelling of the anomalies
between the bosonic and fermionic degrees of freedom [\Ref{AAG}]
and exact S-matrices have been proposed by K\"oberle and Kurak [\Ref{SMCPN}].
The structure of the quantum supersymmetric $\CP$ models is simpler
than that of their bosonic counterparts in that they do not display
confinement [\Ref{INST},\Ref{DDL2}]. Even so they are highly
non-trivial,
asymptotically-free quantum
field theories with dynamically generated mass,
and they have been used---along with their bosonic counterparts---to
gain some profound insights into non-perturbative effects in field theories,
such as the relationship between
instantons and the $1/n$ expansion
(see {\it
eg\/}.~[\Ref{Raj},\Ref{INST},\Ref{DDL2},\Ref{DDL1},\Ref{Col}] and
references therein.)
More recently they have also attracted
attention as massive integrable perturbations of $N=2$
superconformal field theories and as useful examples
to which new non-perturbative techniques can be applied
[\Ref{FI},\Ref{CV},\Ref{BD}].

\chapter{The model and its S-matrix}

The supersymmetric $\CP$ model is defined by a lagrangian density
$$\eqalign{
&{\cal L}={1\over2g} \left \{ | (\del_\mu - A_\mu) z_a |^2
+  i \bar \psi_a \gamma^\mu (\del_\mu - A_\mu) \psi_a \right.\cr
&\qquad\qquad\qquad\left.
+ {1 \over 4} \left[ \left
(\bar \psi_a \psi_a\right)^2  + \left(\bar \psi_a \gamma_5 \psi_a\right)^2
- \left(\bar \psi_a \gamma_\mu \psi_a\right )^2\right]\right\},\cr
&{\rm where} \qquad A_\mu  = \half \left( z^*_a \del_\mu z_a - z_a \del_\mu
z^*_a\right) .
\cr}
\nfr{LAG}
We work throughout in Minkowski space and our conventions agree with
those of [\Ref{REN}].
The fields $z_a$ and $\psi_a$ are an $n$-component complex scalar field and
an $n$-component
complex Dirac fermion, respectively, which satisfy the constraints
$z_a^* z_a^{\phantom{*}} = 1$ and $z_a^* \psi_a^{\phantom{*}} = 0$. The
theory is clearly invariant under
global SU($n$) transformations on these fields. There is also a local U(1)
invariance under which the quantity $A_\mu$ above transforms as a gauge
field; this means that the theory can be interpreted as a sigma-model
whose target manifold is the complex projective space $\CP$.

Applying well-known general results (see {\it eg\/}.~[\Ref{REN}] and
[\Ref{GVZ}]), the
two-loop beta-function for this model and the corresponding behaviour of
the running coupling constant can be written
$$\eqalign{
\beta(g) &= - \beta_1 g^2 - \beta_2 g^3 + {\cal O}(g^4), \cr
&{\rm so} \quad {1 \over g(\mu / \Lambda)} = \beta_1 \ln {\mu \over \Lambda}
+ {\beta_2 \over \beta_1} \ln \ln {\mu \over \Lambda} + {\cal O}
\left ( {\ln \ln (\mu / \Lambda) \over \ln (\mu / \Lambda)} \right ) , \cr
&{\rm where} \quad \beta_1 = n / \pi , \quad \beta_2 = 0 .\cr
}\efr
{}From these expressions we see that the theory is asymptotically free,
with dynamical mass generation.

It is natural to expect the spectrum of the theory to contain
supersymmetric multiplets of particles, some of which are
``fundamental'', in the sense that they carry the quantum
numbers of the fields in the classical lagrangian, and others which can
be regarded as bound states.
We denote the quantum states corresponding to the fundamental particles by
$|a,i,\t\rangle$, where $i=0,1$ denotes a boson or
fermion respectively, $a$ is the $\SU$ vector index of the $n$ dimensional
representation and $\t$ is the rapidity of
the particle, so its velocity is $v={\rm tanh}(\t)$. The
spectrum also includes a set of fundamental
anti-particles $|\bar a,i,\t\rangle$ transforming in the $\bar n$
representation of $\SU$. It turns out that these fundamental
anti-particles can be formed as bound-states of the fundamental
particles (or vice-versa; an example of ``nuclear democracy'').

The integrability of the model [\Ref{AAG}] implies that
the S-matrix factorizes and that all S-matrix elements can be deduced from
the two-body one [\Ref{ZZ}].
On the basis of this, K\"oberle and Kurak
[\Ref{SMCPN}] (see also [\Ref{FI},\Ref{AL},\Ref{AFL},\Ref{AA}])
proposed an S-matrix to describe the
scattering of the fundamental particle multiplet
in accordance with the SU($n$) symmetry of the model
and all the usual axioms of $S$-matrix theory.
Their proposal can be written in an illuminating way in which
the supersymmetric and $\SU$ degrees of freedom are factored.
This was made explicit in [\Ref{FI}] following the discussion for a general
supersymmetric theory in [\Ref{SHOU}].

In detail, the
two-body $S$-matrix elements for the fundamental particles
can be written
$$
\langle c,k,\t_2;d,l,\t_1,{\rm out}|a,i,\t_1;b,j,\t_2,{\rm in}\rangle=
S_{N=2}(\t_1-\t_2)_{ij}^{kl}S_{\rm CGN}(\t_1-\t_2)_{ab}^{cd}.
\nfr{SM}
The $\SU$ part of the S-matrix is the factorizable S-matrix of the
fundamental vector particle of the $\SU$ chiral Gross-Neveu model
[\Ref{SMCGN}]:
$$
S_{\rm CGN}(\t)_{ab}^{cd}=Y_1(\t)\left[\delta^{ad}\delta^{bc}
-{2\pi i\Delta\over\t}\delta^{ac}\delta^{bd}\right],
\efr
with $\Delta=1/n$ and with the unitarizing/crossing scalar factor
$$
Y_1(\t)={\Gamma(1+i\t/2\pi)\Gamma(-\Delta-i\t/2\pi)\over
\Gamma(-i\t/2\pi)\Gamma(1-\Delta+i\t/2\pi)}.
\efr
The supersymmetric part of the S-matrix is the $N=2$ supersymmetric
${\Bbb Z}_n$ minimal S-matrix [\Ref{FI}] which has the form
$$\eqalign{
&S_{N=2}(\t)=Y_2(\t)\cr
&\times\pmatrix{\sinh(\t/2+i\pi\Delta)&0&0&0\cr
0&i\sin(\pi\Delta)&\sinh(\t/2)&0\cr
0&\sinh(\t/2)&i\sin(\pi\Delta)&0\cr
0&0&0&-\sinh(\t/2-i\pi\Delta)\cr}.\cr}
\nfr{SUSYSM}
where the rows and columns are labelled in the order
$(0,0),(0,1),(1,0),(1,1)$ and where
$$\eqalign{
&Y_2(\t)={1\over\sinh(\t/2+i\pi\Delta)}\cr
&\times\prod_{j=1}^\infty
{\Gamma^2(i\t/2\pi+j)\Gamma(-i\t/2\pi+j+\Delta)\Gamma(-i\t/2\pi+j-\Delta)
\over\Gamma^2(-i\t/2\pi+j)\Gamma(i\t/2\pi+j+\Delta)
\Gamma(i\t/2\pi+j-\Delta)}.\cr}
\efr
The S-matrix elements of the fundamental anti-particles can be found
by crossing the elements \SM.

It is important that in \SM\ we have chosen an ordering
in the final state where the particle of rapidity $\t_2$ is to the left
of the particle of rapidity $\t_1$; it is only this ``modified''
S-matrix (using the nomenclature of [\Ref{SHOU}]) which exhibits the
factorization between the
supersymmetric and bosonic degrees of freedom manifested in \SM.
This is opposite to the choice made in the original treatment of
[\Ref{SMCPN}], which means that some care is required in comparing
the signs for amplitudes involving two fermions.
It is also worth pointing out that although it is this same ``modified''
S-matrix which appears in [\Ref{FI}], we are including the simple pole
in the $\SU$ factor rather than in the supersymmetric factor; the net result
is easily seen to agree with [\Ref{FI}].

The expression \SM\ is ``minimal'' in the sense that it has
the minimum number of poles and zeros on the physical strip (the
region $0\leq{\rm Im}(\t)\leq\pi$) consistent with the
requirements of symmetry, existence of a bound-state
and the axioms of S-matrix theory. But this still leaves open
the possibility of adding CDD factors to the S-matrix;
these spoil none of the axioms, they introduce no new poles on the
physical strip and they passively respect the bootstrap equations.
For our model the CDD ambiguities correspond to multiplying the S-matrix of
the fundamental particles \SM\ by factors of the form
$$
{\sinh\left({\t\over2}-{i\pi\over2 n}(2-\alpha)\right)
\sinh\left({\t\over2}-{i\pi\over2 n}\alpha\right)\over
\sinh\left({\t\over2}+{i\pi\over2 n}(2-\alpha)\right)
\sinh\left({\t\over2}+{i\pi\over2 n}\alpha\right)}.
\nfr{CDDF}
where $0<\alpha<2$.
One of the conclusions of this paper will be that the minimal
form \SM\ is the true S-matrix of the theory, so that all CDD factors are
ruled out. As in previous work [\Ref{EH}-\Ref{ACFT}] this will follow from
the consistency of our calculation using the minimal S-matrix with a
calculation in perturbation theory.

Although it will not concern us directly here, we point out that the
complete spectrum of the model can be determined using the bootstrap
procedure.
The S-matrix \SM\ for the fundamental particles
has a simple pole on the physical strip at $\t=2\pi
i\Delta$ which corresponds to a bound state transforming in the
antisymmetric tensor representation of $\SU$. Continuing the
bootstrap in this way one finds a spectrum of bound-states
which is identical to the
$\SU$ chiral Gross-Neveu model, namely $m_r=m\sin(\pi
r\Delta)/\sin(\pi\Delta)$, $1\leq r<n$, where the $r^{\rm th}$
bound-state transforms in the $r^{\rm th}$ fundamental representation
of $\SU$ (each particle carries in addition supersymmetric quantum
numbers). The fundamental anti-particles correspond to the $n{-}1^{\rm th}$
bound-state. We shall only
require the S-matrix elements of the fundamental particles and the
fundamental anti-particles for our calculation.

\chapter{Coupling to a conserved charge}

As summarized in [\Ref{EH}], we wish to couple the model
to a chemical potential $h$ via a conserved charge $Q$ and to calculate
the resulting free energy as a function of $h$. In other words,
we seek the response $\delta f(h)=f(h)-f(0)$ of the ground state energy
density for the system with Hamiltonian modified from $H$ to $H- hQ$.
The TBA equations allow one to find an expansion for $\delta f(h)$
of the form $h^2F_1(h/m)$ valid when $h\gg m$ and in this
regime one can calculate the same quantity in perturbation theory
to obtain an expression $h^2F_2(h/\Lambda)$, where $\Lambda$ is the usual
dimension-full parameter of perturbation theory. Setting
$F_1(h/m)=F_2(h/\Lambda)$ and comparing the first few leading order
terms gives a powerful test of the S-matrix and allows one
to extract the mass gap $m/\Lambda$.

The analysis of the TBA system is only tractable if we can choose $Q$ so
that the new ground-state consists of a restricted number of particle
types. In [\Ref{EH}] we found that the presence of supersymmetry complicated
matters because the ground state formed a supersymmetric doublet with
non-trivial scattering, unlike the examples in
[\Ref{HMN}--\Ref{FNW}] where one could select $Q$ so as to give a single
particle type.
Despite this complication, the problem considered in [\Ref{EH}]
could still be solved.
In the present case we might be tempted
to make the same choice as for the $\SU$ principle
chiral model [\Ref{BNNW}] and the chiral Gross-Neveu model
[\Ref{CGN}]:
$$
Q={\rm diag}\left(1,-{1\over n-1},\ldots,-{1\over n-1}\right),
\nfr{BC}
for which the fundamental multiplet $|1,j,\t\rangle$ has the largest
charge/mass
ratio. As we shall see in the next section, however, this choice is
inconvenient because, even if the TBA system proved tractable,
it would require a three-loop perturbation theory
computation to extract the mass-gap. To avoid this we are motivated to
consider the alternative choice
$$
Q={\rm diag}\left(1,-1,0,\ldots,0\right).
\nfr{GC}
With this coupling the situation is still more complicated than in [\Ref{EH}]
because there are now two fundamental doublets with the largest
charge/mass ratio,
namely
$|1,j,\t\rangle$ and $|\bar2,j,\t\rangle$,
a feature which clearly
arises because of the presence of distinct antiparticles.
Nevertheless we shall see that the resulting TBA system can be analyzed
successfully and that the mass gap can again be extracted by comparison with a
perturbation theory calculation to just one loop.

It is worth re-iterating the point already emphasized in [\Ref{EH}]
that it is an assumption that only
$|1,j,\t\rangle$ and $|\bar2,j,\t\rangle$
appear in the ground state and that in particular
no bound states contribute.
As in most previous work of this type, we rely on the consistency of our
final results to vindicate this assumption.
An important property of the states which we are assuming appear,
is that their scattering is purely elastic in the
space of $\SU$ quantum numbers. Of course the
scattering is still non-diagonal in the supersymmetric subspace. The
$\SU$ part of the S-matrix elements of
1 with 1 and $\bar2$ with $\bar2$ coincide:
$$
S_{\rm CGN}(\t)_{11}^{11}=S_{\rm
CGN}(\t)_{\bar2\bar2}^{\bar2\bar2}=
{\Gamma(1+i\t/2\pi)\Gamma(1-\Delta-i\t/2\pi)\over
\Gamma(1-i\t/2\pi)\Gamma(1-\Delta+i\t/2\pi)}.
\nfr{SSEO}
The $\SU$ part of the S-matrix for $1$ interacting with $\bar2$ can be found by
crossing:
$$
S_{\rm CGN}(\t)_{1\bar2}^{\bar21}=S_{\rm
CGN}(\t)_{\bar21}^{1\bar2}=
{\Gamma(\half-i\t/2\pi)\Gamma(\half-\Delta+i\t/2\pi)\over
\Gamma(\half+i\t/2\pi)\Gamma(\half-\Delta-i\t/2\pi)}.
\nfr{SSET}
A crucial point is that
there is no reflection amplitude for the interaction of
1 and $\bar2$: the scattering is purely elastic.

\chapter{Free-energy from perturbation theory}

To couple the theory to the charge \GC\ by changing the Hamiltonian
$H \rightarrow H- h Q$ we can make the replacement $\del_0 \rightarrow
\del_0 + i h Q$ in the Lagrangian \LAG .
It will turn out to be sufficient to calculate the ground-state energy
density of this theory to one loop.
We must therefore expand the Lagrangian to quadratic order in an independent
set of fields and we can drop all terms which are independent of $h$ to this
order because we are interested only in the response of the
free-energy density,
$\delta f(h) = f(h) - f(0)$. By exploiting the local U(1) invariance of the
action, we can take $z_1$ to be real and we can solve the bosonic
constraint by writing $z_1 = \sqrt{ (1 - | \pi |^2 )(\half + \phi) }$ and
$z_2 = e^{i \theta} \sqrt{ (1 - | \pi |^2)(\half - \phi) }$
where $\pi = (z_3 , \ldots , z_n)$ and $\theta$, $\phi$ are real.
The fermionic degrees of freedom and
the variable $\theta$ decouple to this order, and we are left with the
expression
$$
{\cal L}_{{\rm 1-loop}} = {1 \over 2 g} \left \{ (\del_\mu \phi)^2
+ | \del_\mu \pi |^2 + h^2 - 4 h^2 \phi^2 - h^2 | \pi |^2 \right\}.
\efr

Using dimensional regularization with the $\overline{\rm MS}$-scheme gives
the one-loop free energy
$$
\delta f (h) =  -{h^2 \over 2 g} - {h^2 \over \pi} \ln 2
+ {n h^2 \over 4 \pi} (1 - \ln (h^2 / \mu^2)).
\efr
We now substitute for the running coupling.
The result, expressed in terms of the one-loop and
two-loop beta functions coefficients, is:
$$\eqalign{
&\delta f(h)=\cr
&-h^2 {\beta_1\over2}\left[\ln{h\over
\Lambda_{\overline{\rm MS}}}-{1\over2}+{2 \ln 2 \over \pi \beta_1}
+{\beta_2 \over \beta_1^2} \ln \ln {h \over \Lambda_{\overline{\rm MS}} }
+{\cal O}\left
({\ln\ln(h/\Lambda_{\overline{\rm
MS}})\over\ln(h/\Lambda_{\overline{\rm MS}})}\right)\right].\cr}
\nfr{FEP}
Substituting the specific values for these coefficients given in (2.2)
we obtain
$$
\delta f(h)=-{h^2n\over2\pi}\left[\ln{h\over
\Lambda_{\overline{\rm MS}}}-{1\over2}+{2 \ln 2 \over n}+{\cal O}\left
({\ln\ln(h/\Lambda_{\overline{\rm
MS}})\over\ln(h/\Lambda_{\overline{\rm MS}})}\right)\right].
\nfr{FEP}
Notice that if we chose the charge $Q$ to be \BC\ then there would be
no ``tree-level'' ${\cal O}(1/g)$ term. In this case it would be
necessary to do a three-loop computation in order to extract the mass-gap.

\chapter{Free-energy from the S-matrix}

In this section we write down the TBA equations for the model and
solve them in the limit $h\gg m$.
We must follow the hypothesis made earlier
that only the multiplets
$|1,j,\t\rangle$ and $|\bar2,j,\t\rangle$ contribute to the
ground-state. Since the scattering of these multiplets is purely
elastic, it will not be necessary to perform a diagonalization in the
space of $\SU$ quantum numbers (although this diagonalization can be
done [\Ref{THII}]). The remaining difficulty is that the S-matrix
for these favoured states is still
non-diagonal in the supersymmetric subspace. But this problem can also
be solved since it has been shown by
Fendley and Intriligator [\Ref{FI}] that it is equivalent to
diagonalizing the transfer matrix of the six vertex model at the free fermion
point.

In [\Ref{FI}] a system of equations is derived which links the
density of single particle states in rapidity space $\varrho_a(\t)$ to the
density of occupied single particle states $\sigma_a(\t)$, where $a=1,\bar2$.
The equations also involve densities $P^+_l(\t)$ and $P^-_l(\t)$, $l=0,\bar0$
corresponding to two ``supersymmetric magnons'', reflecting the
fact that there are two supersymmetries:
$$\eqalign{
&\varrho_a(\t)={m\over2\pi}\cosh\t+\phi_{ab}*\sigma_b(\t)+
\phi_{al}*P^+_l(\t),\cr
&P^+_l(\t)+P^-_l(\t)=\phi_{al}*\sigma_a(\t),\cr}
\nfr{BA}
where $f*g(\t)=\int_{-\infty}^\infty
d\t'f(\t-\t')g(\t')$. The kernels appearing here are
$$
\eqalign{
\phi_{ab}(\t)&={1\over2\pi i}{d\over d\t}
\ln S_{\rm CGN}(\t)_{ab}^{ba},\cr
\phi_{10}(\t)&=\phi_{\bar2\bar0}(\t)=
{\sin(\pi\Delta)\over\cosh\t-\cos(\pi\Delta)},\cr
\phi_{\bar20}(\t)&=\phi_{1\bar0}(\t)=
{\sin(\pi\Delta)\over\cosh\t+\cos(\pi\Delta)},
\cr}
\efr
where as before $a,b=1,\bar2$ and the elements $S_{\rm
CGN}(\t)_{ab}^{ba}$ are written down in \SSEO\ and \SSET.
Notice that the kernel which multiplies the densities of
the particles $\sigma_a(\t)$ involves
a contribution only from the chiral Gross-Neveu part of the S-matrix---so
if we removed the magnon terms the equations would be identical to
those for the chiral Gross-Neveu model. Notice also that the
super-magnons do not interact amongst themselves.

To find the TBA equations [\Ref{TBA}]
we define the ``excitation energies'' of the particles
$\epsilon_a(\t)$ and the magnons $\xi_l(\t)$ at finite temperature $T$ by
$$
{\sigma_a(\t)\over\varrho_a(\t)}={1\over e^{\epsilon_a(\t)/T}+1},\qquad
{P^-_l(\t)\over P^+_l(\t)}=e^{\xi_l(\t)/T}.
\efr
We shall require the TBA equations at zero temperature with the field $h$
coupling via $Q$ acting as a chemical potential.
At $T=0$,
$\epsilon_a(\pm\t_{\rm F}^a)=0$, where
$\t_{\rm F}^a$ are Fermi rapidities and $\epsilon_a(\t)$ is negative
precisely when
$-\t_{\rm F}^a<\t<\t_{\rm F}^a$.
It is convenient to introduce the
following notation
$$
f^\pm(\t)=\cases{f(\t)\qquad&$f(\t){>\atop<}0$\cr 0&otherwise.\cr}
\efr
The free-energy per unit volume at $T=0$
is
$$
\delta f(h)=f(h)-f(0)={m\over2\pi}\int_{-\infty}^\infty d\t
\left[\epsilon_1^-(\t)+\epsilon_{\bar2}^-(\t)\right]\cosh\t,
\nfr{FE}
where $\epsilon_a(\t)$, $a=1,\bar2$, are the solutions of the
$T=0$ TBA equations:
$$
\eqalign{
\epsilon_a(\t)-\phi_{ab}*\epsilon^-_b(\t)-\phi_{al}*\xi^-_l(\t)&=
m\cosh\t-h,\cr
\xi_l(\t)-\phi_{al}*\epsilon^-_a(\t)&=0.\cr}
\nfr{TBA}

To simplify \TBA\ it is important to notice that $\phi_{al}(\t)$ is a
positive kernel, which implies that the magnon variables are given by
$\xi^+_l(\t)=0$ and $\xi^-_l(t)=\phi_{al}*\epsilon^-_a(\t)$.
Furthermore, the solution does
not distinguish between the values of the favoured SU($n$) quantum numbers
and so we have $\epsilon_1(\t)=\epsilon_{\bar2}(\t)\equiv\epsilon(\t)$.
The four equations in \TBA\ then reduce
to a single equation for $\epsilon(\t)$:
$$
\epsilon^+(\t)+R*
\epsilon^-(\t)=m\cosh\t-h,
\nfr{RTBA}
where the kernel is
$$
R(\t)=\delta(\t)-\phi_{11}(\t)-\phi_{1\bar2}(\t)-\left[\phi_{10}+\phi_{1\bar0}
\right]*\left[\phi_{10}+\phi_{1\bar0}\right](\t).
\efr
and the expression \FE\ for the free-energy density becomes
$$
\delta f(h)={m\over\pi}\int_{-\infty}^\infty d\t\,\epsilon^-(\t)\cosh\t.
\efr

The Fourier transform of the kernel in \RTBA\ is
$$
R(\t)
=\int_0^\infty{d\omega\over\pi}\cos(\omega\t){\cosh((1-2\Delta)\pi
\omega/2)\sinh(\pi\Delta\omega)\over\cosh^2(\pi\omega/2)}e^{\pi\omega/2},
\efr
where $\Delta=1/n$.
As in [\Ref{EH}] we see that this
Fourier transform vanishes at the origin so that the solution
resembles those for the bosonic models discussed in
[\Ref{HMN}-\Ref{BNNW}].
We now seek an expression for the Fourier transform of the kernel
$1/(G_+(\omega)G_-(\omega))$ where $G_\pm(\omega)$ are analytic
in the upper (lower) half planes and $G_-(\omega)=G_+(-\omega)$.
The unique solution is
$$\eqalign{
G_+(\omega)=&{\Gamma(\half-i(1-2\Delta)\omega/2)\Gamma(1-i\Delta\omega)\over
\Gamma^2(\half-i\omega/2)}e^{-\half\ln(-i\Delta\omega)}\cr
&\times e^{-i\omega(\half-\Delta)(1-\ln(-i\omega(\half-\Delta)))
-i\omega\Delta(1-\ln(-i\omega\Delta))+i\omega(1-\ln(-i\half\omega))}.\cr}
\efr
{}From [\Ref{BNNW}] we know that if
$G_+(i\xi)$ has an expansion for small $\xi$ like
$$
G_+(i\xi)={k\over\sqrt\xi}e^{-a\xi\ln\xi}\left(1-b\xi+{\cal
O}(\xi^2)\right),
\nfr{EXG}
then the free-energy density for $h\gg m$ takes the form
$$\eqalign{
\delta f(h)=&-{h^2k^2\over2}\left[\ln{h\over m}
+\ln\left({\sqrt{2\pi}ke^{-b}\over G_+(i)}\right)-1+a(\gamma_{\rm E}
-1+\ln8)\right.\cr
&\qquad\qquad\qquad\left.+(a+\half)\ln\ln{h\over m}+
{\cal O}\left({\ln\ln(h/m)\over\ln(h/m)}\right)\right].\cr}
\efr
Our kernel does indeed have an expansion of the form \EXG\ with
$$
k={1\over\sqrt{\pi\Delta}},\quad
a=-{1\over2},\quad{\sqrt{2\pi}ke^{-b}\over G_+(i)}={\sin(\pi\Delta)
\over\pi\Delta}e^{\gamma_{\rm E}/2+\left(\frac{3}{2}+2\Delta\right)\ln2}.
\efr
The resulting expression for the free energy is
$$
\delta f(h)=-{h^2\over2\pi\Delta}\left[\ln{h\over m}
+\ln\left({\sin(\pi\Delta)\over\pi\Delta}2^{2\Delta}\right)-{1\over2}+
{\cal O}\left({\ln\ln(h/m)\over\ln(h/m)}\right)\right].
\nfr{FESM}

\chapter{Comparison and conclusions}

Comparing \FESM\ with \FEP\ we see that the result from the TBA
calculation correctly reproduces the universal coefficients of the
beta-function and we extract the following value for the mass-gap for
the supersymmetric $\CP$ model
$$
{m\over\Lambda_{\overline{\rm MS}}}={\sin(\pi\Delta)\over
\pi\Delta},\qquad\Delta={1\over n}, \qquad n > 1.
\nfr{MG}

It is instructive to consider what would have happened if we had
chosen the other charge \BC , rather than \GC . In that case only the
multiplet
$|1,j,\t\rangle$ would have appeared in the ground-state and the
resulting TBA equation would have had the same form as \RTBA, but with
a different kernel:
$$
R(\t)=\delta(\t)-\phi_{11}(\t)-\phi_{10}*\phi_{10}(\t)-\phi_{1\bar0}*
\phi_{1\bar0}(\t).
\efr
The Fourier transform of this kernel does not vanish at the origin,
and so the situation is analogous to the Gross-Neveu
models [\Ref{CGN},\Ref{FNW}]. In these cases, the expansion of the
free-energy density for $h\gg m$ has a different form which has a
well-defined limit as $h\rightarrow\infty$, unlike \FESM.
This matches precisely the
fact that with this different choice of charge the perturbative
expansion of the free-energy density is also markedly different since
there is no ``tree-level'' contribution and it would require a
three-loop calculation to extract the mass-gap.

In addition to our result for the mass gap, an important conclusion
of our paper is that we have resolved the problem of CDD ambiguities in the
ansatz of K\"oberle and Kurak.
Any additional CDD factors of the form
\CDDF\ would alter the kernel $R(\t)$ of the TBA equation
and the thermodynamics of the system depends so sensitively on this
that we can argue with some confidence that the
solution for the free-energy density would not match that found in
perturbation theory.

It is interesting to compare our result for the mass-gap
with the spectrum for the super-$\CP$ model proposed in [\Ref{CV}]
on the basis of very different methods.
It is clear that the the $n$-dependence of these expressions
is in complete agreement (setting $r = 1$ in the formula given in
the discussion following equation (6) in the first reference in
[\Ref{CV}] should give the mass of the fundamental multiplet)
but the connection between the overall scale
factors is less clear. It would be interesting to investigate in
detail how the results of [\Ref{CV}] could be expressed in terms of
the more conventional scale parameter used in this paper.

J.M.E. is grateful to Michele Bourdeau for discussions concerning
[\Ref{FI}-\Ref{BD}].

\references

\beginref
\Rref{FNW}{P. Forg\'acs, F. Niedermayer and P. Weisz, Nucl. Phys. {\bf
B367} (1991) 123}
\Rref{HN}{P. Hasenfratz and F. Niedermayer, Phys. Lett. {\bf B245}
(1990) 529}
\Rref{BNNW}{J. Balog, S. Naik, F. Niedermayer and P. Weisz, Phys. Rev. Lett.
{\bf69} (1992) 873\newline
S. Naik, Nucl. Phys. {\bf B} (Proc. Suppl.) {\bf30} (1993) 232}
\Rref{HMN}{P. Hasenfratz, M. Maggiore and F. Niedermayer, Phys. Lett.
{\bf B245} (1990) 522}
\Rref{ZZ}{A.B. Zamolodchikov and Al. B. Zamolodchikov, Ann. Phys.
{\bf120} (1979) 253}
\Rref{THII}{T.J. Hollowood, Phys. Lett. {\bf B320} (1994) 43}
\Rref{THIII}{T.J. Hollowood, Phys. Lett. {\bf B329} (1994) 450}
\Rref{TBA}{E.H. Lieb and W. Liniger, Phys. Rev. {\bf130} (1963) 1605\newline
Al.B. Zamolodchikov, Nucl. Phys. {\bf B342} (1990) 695}
\Rref{SMCGN}{B. Berg and P. Weisz, Nucl. Phys. {\bf B146} (1979)
205\newline
V. Kurak and J.A. Swieca, Phys. Lett. {\bf B82} (1979) 289}
\Rref{CGN}{P. Forgacs, S. Naik and F. Niedermayor,
Phys. Lett. {\bf B283} (1992) 282}
\Rref{REN}{L. Alvarez-Gaum\'e, D.Z. Freedman and S.K. Mukhi, Ann.
Phys. {\bf134} (1981) 85}
\Rref{GVZ}{M.T. Grisaru, A.E.M. de Ven and D.
Zanon, Nucl. Phys. {\bf B277}
(1986) 388; 409}
\Rref{SHOU}{K. Schoutens, Nucl. Phys. {\bf B344} (1990) 665}
\Rref{ACFT}{J.M. Evans and T.J. Hollowood, {\sl Integrable theories that
are asymptotically CFT\/}, Preprint CERN-TH.7293/94, SWAT/93-94/32,
{\tt hep-th/9407113}}
\Rref{FI}{P. Fendley and K. Intriligator, Nucl. Phys. {\bf B380}
(1992) 265}
\Rref{AL}{E. Abdalla and A. Lima-Santos, Phys. Rev. {\bf D29}
(1984) 1851}
\Rref{SMCPN}{R. K\"oberle and V. Kurak, Phys. Rev. {\bf D36} (1987) 627}
\Rref{AFL}{E. Abdalla, M. Forger and A. Lima Santos, Nucl. Phys. {\bf
B256} (1985) 145}
\Rref{AA}{E. Abdalla and M.C.B. Abdalla, {\sl Exact S matrices and
extended supersymmetry\/}, {\tt hep-th/9311178}}
\Rref{EH}{J.M. Evans and T.J. Hollowoood, {\sl The exact mass-gap of
the supersymmetric O(N) sigma model\/}, CERN-TH.7425/94,
SWAT/93-94/30}
\Rref{AAG}{E. Abdalla, M.C.B. Abdalla and M. Gomes, Phys. Rev. {\bf
D25} (1982) 452; Phys. Rev. {\bf D27} (1983) 825}
\Rref{CV}{S. Cecotti and C. Vafa, Phys.~Rev.~Lett.~{\bf 68} (1992) 903;
Commun.~Math.~Phys.~{\bf 158} (1993) 569}
\Rref{BD}{M. Bourdeau and M. Douglas, Nucl.~Phys.~{\bf B420} (1994) 243}
\Rref{Col}{S. Coleman, ``1/N'' in ``Aspects of Symmetry'' (C.U.P., 1984)}
\Rref{INST}{E. Witten, Nucl.~Phys.~{\bf B149} (1979) 285}
\Rref{CS}{E. Cremmer and S. Scherk, Phys.~Lett.~{\bf B74} (1978) 341}
\Rref{DDL1}{A. D'Adda, P. Di Vecchia and M. L\"uscher, Nucl.~Phys.~{\bf B146}
(1978) 63}
\Rref{DDL2}{A. D'Adda, P. Di Vecchia and M. L\"uscher, Nucl.~Phys.~{\bf B152}
(1979) 125}
\Rref{Raj}{R. Rajaraman, ``Solitons and Instantons'' (North-Holland, 1987)}
\endref
\ciao